\documentclass[12pt]{article}
\usepackage{pdproc} 
\usepackage{graphicx} 

\newcommand{\be}{\begin{equation}}
\newcommand{\ee}{\end{equation}}
\newcommand{\bea}{\begin{eqnarray}}
\newcommand{\eea}{\end{eqnarray}}

\newcommand{\AmS}{{\protect\the\textfont2
  A\kern-.1667em\lower.5ex\h\citebox{M}\kern-.125emS}}
\newcommand{\lsim}{\mathrel{\mathop{\kern 0pt \rlap
  {\raise.2ex\hbox{$<$}}}
  \lower.9ex\hbox{\kern-.190em $\sim$}}}
\newcommand{\gsim}{\mathrel{\mathop{\kern 0pt \rlap
  {\raise.2ex\hbox{$>$}}}
  \lower.9ex\hbox{\kern-.190em $\sim$}}}
\def\Journal#1#2#3#4{{#1} {{\bf #2}} (#3) #4 }

\def\APP{\em Astrop. Phys.}

\def\NIMA{{\em Nucl. Instr. \& Meth.} {\bf A}}

\def\PLB{{\em Phys. Lett.} {\bf B}}

\def\PRL{\em Phys. Rev. Lett.}

\def\PRD{{\em Phys. Rev.} {\bf D}}

\begin{document}
\renewcommand{\thefootnote}{\alph{footnote}}

\begin{flushright}
{\bf ROM2F/2003/9} \\
May 2003 \\
\small
Contributed paper to X International Workshop \\
on "Neutrino Telescopes", Venice 2003
\normalsize
\end{flushright}

\vspace{0.5cm}

\title{DAMA RESULTS}

\author{ R. BERNABEI, P. BELLI, F. CAPPELLA, R. CERULLI, 
F. MONTECCHIA\footnote{also: 
Universita' "Campus Biomedico" di Roma, 00155, Rome, Italy}, F. NOZZOLI}

\address{ Dip. di Fisica, Universita' di Roma ''Tor Vergata" \\
and INFN, sez. Roma2, I-00133 Rome, Italy}

\author{A. INCICCHITTI, D. PROSPERI}

\address{Dip. di Fisica, Universita' di Roma ''La Sapienza" \\
and INFN, sez. Roma, I-00185 Rome, Italy}

\centerline{\footnotesize and}

\author{C.J. DAI, H.H. KUANG, J.M. MA, Z.P. YE\footnote{also:
University of Zhao Qing, Guang Dong, China}}

\address{IHEP, Chinese Academy, P.O. Box 918/3, Beijing 100039, China}

\abstract{DAMA is an observatory for rare 
processes based on the development and use
of various kinds of radiopure scintillators. 
Several low background set-ups have
been realized with time passing and many rare processes have been investigated.
Main activities are briefly summarized in the following and the main arguments
on the the results achieved in the investigation of the WIMP annual modulation
signature are addressed. Next perspectives are also mentioned.}

\normalsize\baselineskip=15pt

\section{Introduction}

DAMA is an observatory for rare processes based on the development and use of
various kinds of radiopure scintillators.
The main experimental set-ups are: i) the  $\simeq$ 
100 kg NaI(Tl) set-up, which has completed its data taking in July 2002; ii) the
new 250 kg NaI(Tl) LIBRA (Large sodium Iodide Bulk for RAre processes) set-up,
whose installation is started at fall 2002; iii) the $\simeq$ 6.5 kg liquid Xenon
(LXe) pure scintillator; iv) the R\&D installation for tests on prototypes and small
scale experiments. Moreover, in the framework of devoted R\&D for higher radiopure
detectors and PMTs, sample measurements are regularly carried out by means of
the low background DAMA/Ge detector, installed deep underground since about a
decade and, in some cases, at Ispra.

\begin{figure}[!ht]
\centering
\includegraphics[height=6.5cm]{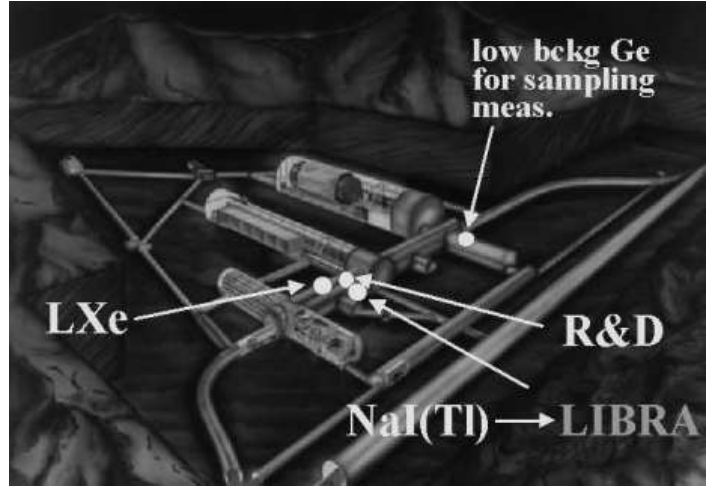}
\caption{The locations of the DAMA experimental installations in the Gran Sasso underground
laboratory of I.N.F.N.}
\label{fg:fig1}
\vspace{-1.0cm}
\end{figure}

The locations of the DAMA experimental installations in the Gran Sasso 
underground laboratory of I.N.F.N. are shown in Fig. \ref{fg:fig1}.

\section{DAMA/LXe}

The DAMA/LXe experiment has followed the former Xelidon experimen on R\&D
developments of liquid Xenon (LXe) detectors and has realized since '90 several LXe
scintillator prototypes using natural Xenon. Then, it has preliminarily put in
measurement the set-up used in the data taking of ref. \cite{1,2} 
by using Kr-free Xenon enriched
in $^{129}$Xe at 99.5\%. This set-up was significantly upgraded at fall 1995 (as mentioned
e.g. in ref. \cite{3}) by including: i) a new purification system without Oxisorb;
ii) a new
low background Pb in the shield; iii) a new low background Cu in the shield and in
the insulation vessel; iv) a substitution of some PMTs and improved low background
voltage dividers, etc. As a consequence, a significant improvement in the counting
rate (as tipically experienced in low background experiments; see e.g. the case of ref.
\cite{4} for Ge detector) was obtained. The rate measured after this upgrading was well
consistent over the whole energy spectrum and during several years of data taking,
although some other minor substitutions have obviously been carried out with time
passing.

In summer 2000 the set-up was again deeply modified (reaching the configuration
reported in Fig.5 of ref. \cite{5}) to handle also Kr-free Xenon enriched in $^{136}$Xe 
at 68.8\%
and in $^{134}$Xe at 17.1\% \cite{6,7}. The main differences among the previous and the present
experimental set-up operating with this latter gas are: the gas\footnote{It was produced by another 
factory and many years before than the one enriched in $^{129}$Xe; in
addition, it was also previously used in a different underground experiment where different 
materials and vacuum/purification/filling/recovery system were operative and, afterwards, stored 
underground for long time in bottles with different possible effects from surface degassing, etc.;
thus, the background is significantly different.}, part of the set-up\footnote{
The vacuum/purification/filling/recovery system has been significantly modified to allow the
allocation and handling of the Xenon enriched in $^{136}$Xe and in some other parts such as the cold
trap (a new concept one) and the shield.} and the isotope\footnote{It can also be interested by 
different physical processes than the $^{129}$Xe previously used.}.
In particular, the high energy rate measured with this new set-up 
has been found 
higher than the one previously measured with the $^{129}$Xe set-up.

\begin{figure}[!ht]
\centering
\includegraphics[height=6.cm]{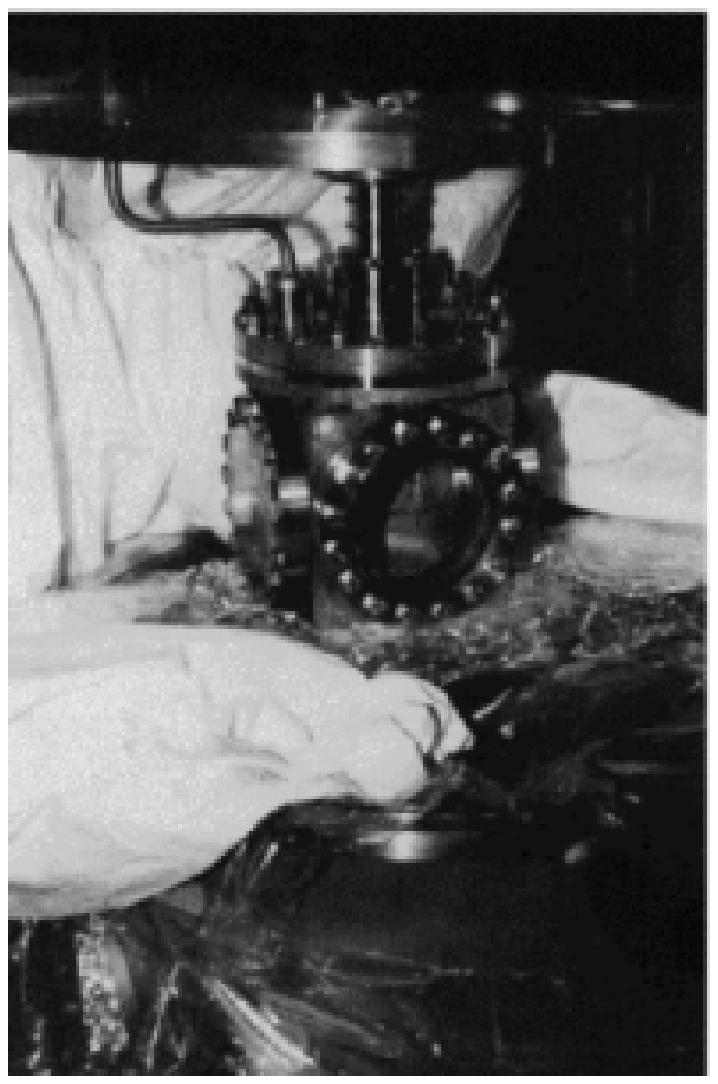}
\includegraphics[height=6.cm]{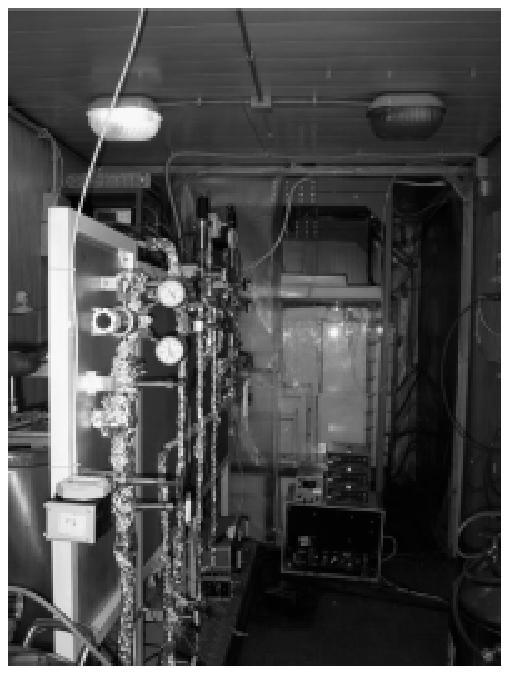}
\caption{On the left: the vessel of the LXe set-up. On the right:
the vacuum/purification/fil\-ling/re\-covery system and the passive shield ahead.}
\label{fg:fig2}
\vspace{-0.2cm}
\end{figure}

The main features of the set-up (see Fig. \ref{fg:fig2}) are described in ref.
\cite{5} and in related papers. 
We also remind that careful neutron calibrations have been carried out during several
years \cite{9,10}.

As regards the more recent results achieved with this set-up on the Dark Matter
investigation we mention the limits on recoils obtained by investigating the 
WIMP-$^{129}$Xe elastic scattering exploiting the pulse shape discrimination technique \cite{9} and
those obtained -- in a given model framework -- on the WIMP-$^{129}$Xe inelastic scattering
\cite{3,11}.

The same experiment has allowed to investigate several other rare processes such
as nuclear level excitation of $^{129}$Xe during charge-non-conserving processes \cite{12}
and the possible electron decay through the channel: $e^- \rightarrow \nu_e + \gamma$ \cite{13}.
In addition, the
nucleon and di-nucleon decay into invisible channels has been investigated with a
new approach \cite{14} based on the search for the radioactive daughter nuclei, created
after the nucleon or di-nucleon disappearance in the parent nuclei. The advantage of
this approach is a branching ratio close to 1 and an efficiency -- since the parent and
the daughter nuclei are located in the detector itself -- also close to 1.
Competitive limits have been obtained.

After the latest upgrading of the set-up double beta decay modes in $^{136}$Xe and
in $^{134}$Xe have been deeply investigated reaching competitive limits as well \cite{6,7}.

The data taking is continuing.

\section{DAMA/R\&D}

\begin{figure}[!htbp]
\begin{center}
\includegraphics[height=5.cm]{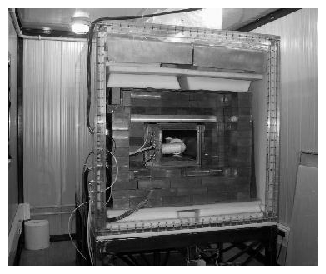}
\end{center}
\vspace{-0.4cm}
\caption{View of the open shield of the R\&D installation.}
\label{fg:fig3}
\vspace{-0.2cm}
\end{figure}

The set-up named "R\&D" is used for tests on prototypes and small scale 
experiments. A view of the passive shield of this installation is given 
in Fig. \ref{fg:fig3}.  

This set-up which has been deeply upgraded in 2000/2001 has been used for measurements
on low background prototype scintillators and PMTs realized in various R\&D efforts
with industries. Moreover, it is regularly used to perform small scale experiments
mainly investigating double beta decay modes in various isotopes.
Among the obtained results we remind the search for: i) $\beta\beta$ 
decay modes in $^{136}$Ce and in $^{142}$Ce \cite{15};
ii) $2EC2\nu$ decay mode in $^{40}$Ca \cite{16};
iii) $\beta\beta$ decay modes in $^{46}$Ca and in $^{40}$Ca \cite{17};
iv)  $\beta\beta$ decay modes in $^{106}$Cd \cite{18};
v) $\beta\beta$ and $\beta$ decay modes in $^{48}$Ca \cite{19};
vi) $2EC2\nu$ in $^{136}$Ce and in $^{138}$Ce and $\alpha$ decay in $^{142}$Ce 
\cite{20}.

\begin{figure}[!ht]
\begin{center}
\vspace{-0.7cm}
\includegraphics[height=9.cm,angle=270]{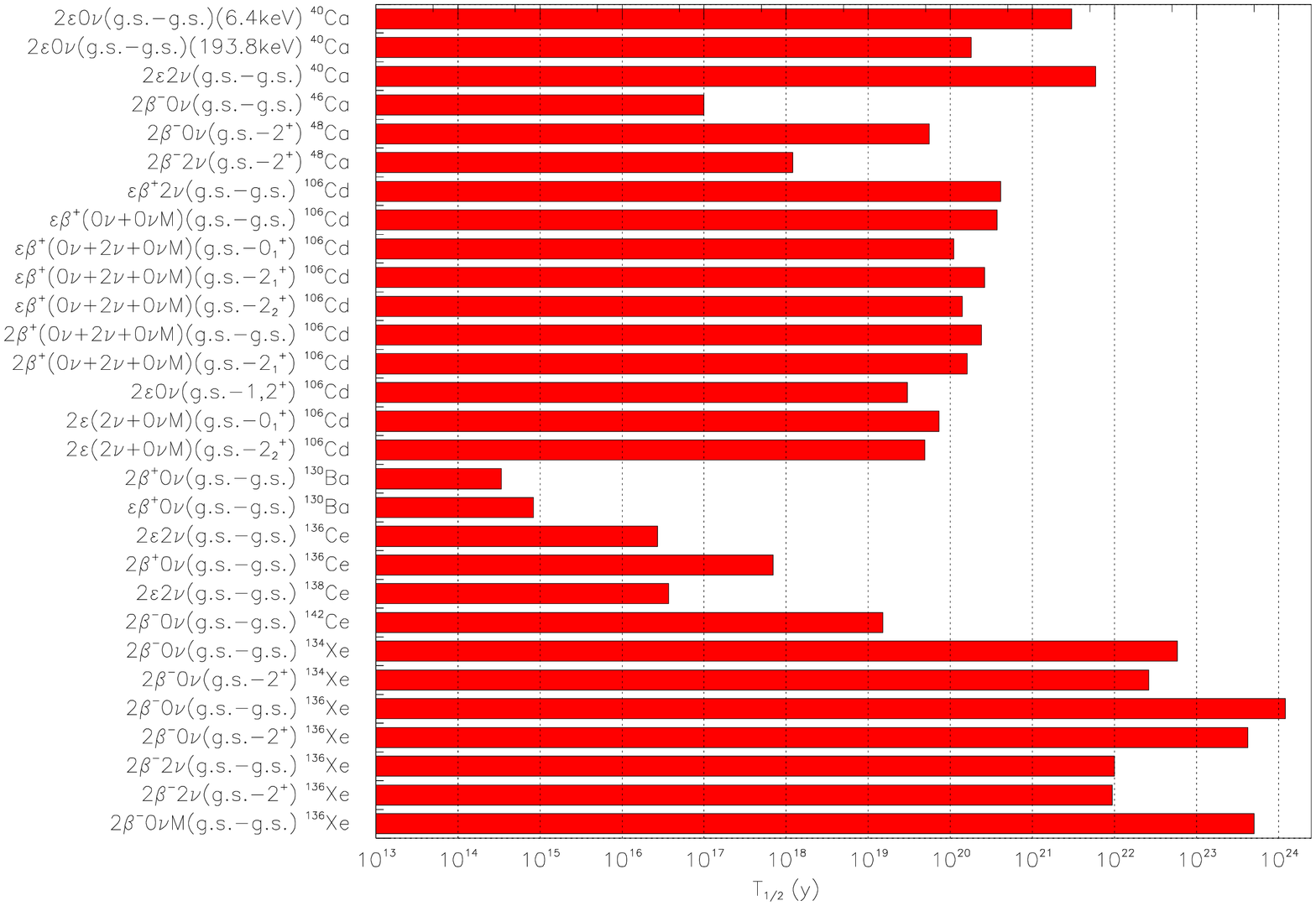}
\end{center}
\vspace{-0.4cm}
\caption{Summary of the limits obtained by DAMA on various double beta decay processes. See
text.}
\label{fg:fig4}
\vspace{-0.2cm}
\end{figure}

Fig. \ref{fg:fig4} summarizes the results obtained
in the searches for double beta decay modes.

\section{DAMA/Ge for sampling measurements}

\noindent Various R\&D developments to
improve low background set-ups and scintillators as well as 
new developments for higher radiopure PMTs are regularly carried out.
The related measurements on samples are performed by means of  
the DAMA low background Ge detector\footnote{This detector was specially realized with a 
low Z window to be sensitive to external radiation down to about ten keV.}, 
which is operative deep underground in the low 
background facility of the Gran Sasso National Laboratory since many years.

\section{DAMA/NaI}

The main goal of the $\simeq$ 100 kg NaI(Tl) (DAMA/NaI)
set-up has been the search for WIMPs by the annual modulation signature,
which it has investigated over seven annual cycles;
the data of 4 of them (about 60000 kg$\cdot$d) have been
already released
\cite{21,22,23,24,25,26,27,28,29} and main arguments will be discussed in the 
following.

\begin{figure}[ht]
\vspace{-0.7cm}
\centering
\includegraphics[height=5.5cm]{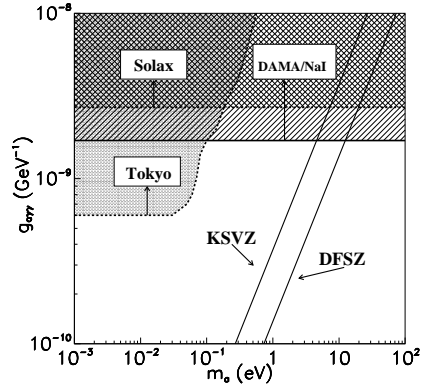}
\caption{Exclusion plot in the plane axion to photon coupling constant,
$g_{a \gamma \gamma}$, versus axion mass, $m_a$, achieved by DAMA/NaI in ref. $^{33)}$.
The limit quoted in the paper ($g_{a \gamma \gamma}\le 1.7 \times 10^{-9}
GeV^{-1}$ at 90\% C.L.) is shown together with
the expectations of the KSVZ and DFSZ models; see ref. $^{33)}$ for details.}
\label{fg:fig5}
\end{figure}

For completeness, we remind that -- in addition to the investigation of the WIMP 
component in the galactic halo 
by means of the annual modulation
signature -- other approaches have also been exploited with DAMA/NaI such as 
the pulse shape discrimination technique \cite{30}
and the investigation of possible diurnal effects \cite{31}.
Moreover, also exotic Dark Matter candidates such as neutral SIMPs,
neutral nuclearities and Q-balls \cite{32,33}
have been searched for. A devoted search for solar axions has been carried out as well 
(see Fig. \ref{fg:fig5}) \cite{34}.

In addition, DAMA/NaI has also allowed to investigate several other rare
processes such as e.g. Pauli exclusion principle violation with spontaneous emission
of protons in $^{23}$Na and $^{127}$I \cite{35},
nuclear level excitation of $^{127}$I 
and $^{23}$Na during charge-non-conserving processes \cite{36} and
electron stability and non-paulian transitions in Iodine atoms (by L-shell)
\cite{37}.

\noindent The set-up and its performances have been described in
details in ref. \cite{22}; since then some upgrading has been
carried out. In particular, in summer 2000 the electronic chain and
data acquisition system have been completely substituted,
while during August 2001 the new HV power supply
system and the new
preamplifiers prepared for the foregoing LIBRA set-up have been
installed here.

\noindent The set-up has completed its data taking in July 2002.
Various kind of data analyses are continuing; in particular,
the total statistics of 107731 kg$\cdot$d
will be presented in near future giving further results 
on the WIMP annual modulation signature.

\subsection{ Results on the WIMP annual modulation signature}

DAMA/NaI has been proposed in '90 \cite{38} and has been realized to investigate with 
suitable sensitivity
the WIMP annual modulation signature, originally proposed in
ref. \cite{39}. It has been so far the only experiment able to test 
this model independent signature for WIMPs with suitable mass, sensitivity and control
of the running parameters.

This WIMP model independent signature is based on the annual modulation of
the signal rate induced by the Earth revolution
around the Sun; as a consequence, the Earth will be 
crossed by a larger WIMP flux roughly in June (when its rotational velocity is summed 
to the one of the solar system with respect to the Galaxy)
and by a smaller one roughly in December (when the two velocities are subtracted).
The fractional difference between the maximum and the minimum of the rate is
of order of $\simeq$ 7\%. Therefore, to point out the modulated component
of the signal, large mass apparata with suitable performances and
control of the operating conditions -- such as the
$\simeq$ 100 kg highly radiopure NaI(Tl) DAMA set-up -- are necessary.

The annual modulation signature is very distinctive
since a WIMP-induced seasonal effect must simultaneously satisfy
all the following requirements: the rate must contain a component
modulated according to a cosine function (1) with one year period (2)
and a phase that peaks around $\simeq$ 2$^{nd}$ June (3);
this modulation must only be found
in a well-defined low energy range, where WIMP induced recoils
can be present  (4); it must apply to those events in
which just one detector of many actually "fires", since
the WIMP multi-scattering probability is negligible (5); the modulation
amplitude in the region of maximal sensitivity must be $\lsim$7$\%$ (6).
Only systematic effects able to fulfil these 6 requirements could fake
this signature; no one has been found nor suggested by anyone 
after several years of investigations.

\subsubsection{The model independent evidence by DAMA/NaI}

\begin{figure}[ht]
\centering
\vspace{-1.0cm}
\includegraphics[height=6.cm]{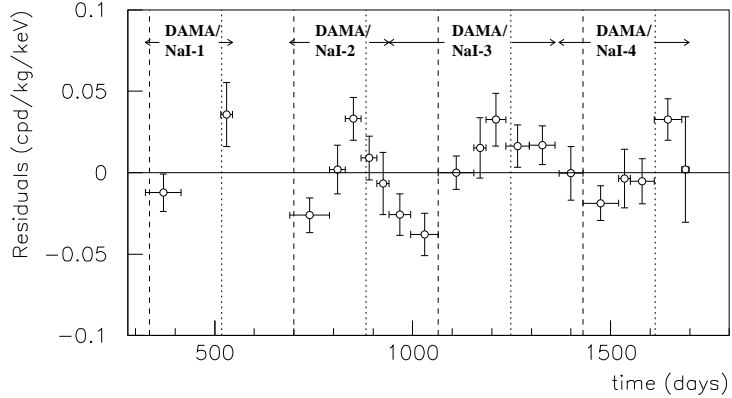}
\caption{Model independent residual rate for single hit events, in the
2--6 keV cumulative energy interval, as a function of the time elapsed
since
January 1-st of the first year of data taking. The expected
behaviour of a WIMP signal is
a cosine function with minimum roughly at the dashed
vertical lines and with maximum roughly at the dotted ones.}
\label{fg:fig6}
\end{figure}

A model independent analysis of the data 
of the four annual cycles has offered an immediate evidence of the presence of
an annual modulation of the rate of the single hit events
in the lowest energy interval (2 -- 6 keV) as shown in Fig. \ref{fg:fig6}.
The $\chi^2$ test on the data of Fig. \ref{fg:fig6}
disfavors the hypothesis of unmodulated behaviour
(probability: $4 \cdot 10^{-4}$), while
fitting these residuals with the function
$A \cdot$ cos$\omega (t-t_0)$, one gets:
i) for the period T = $\frac{2\pi}{\omega}$ = (1.00 $\pm$ 0.01) year
when $t_0$ is fixed at the 152.5$^{th}$ day of the year (corresponding to
$\simeq$ 2 June); ii)
for the phase t$_0$ = (144 $\pm$ 13) days,
when T is fixed at 1 year. In the two cases $A$ is:
(0.022 $\pm$ 0.005) cpd/kg/keV and
(0.023 $\pm$ 0.005) cpd/kg/keV, respectively.
Similar results, but with slightly larger
errors, are found in case all the parameters are kept free.

The modulation is absent in other part of the energy spectrum as discussed 
e.g. in ref. \cite{26}; for example, in the 6-10 keV energy region
just above a modulation amplitude $A = -(0.0017 \pm 0.0037)$ cpd/kg/keV is
obtained.

We have extensively discussed the results of the investigations
of all sources of possible systematics when releasing
the data of each annual cycle; moreover, a dedicated
paper \cite{26} has been devoted to this subject. 
No 
systematic effect or side reaction able to mimic a WIMP induced
effect (that is to be both quantitative relevant and able to satisfy the
six peculiarities of the WIMP signature) 
has been found. This deep discussion will not be repeated here since 
the reader can directly find it in ref. \cite{26} with all the details.

In conclusion, a WIMP contribution
to the measured rate has been candidate by the result of the model independent
approach independently on the nature and coupling
with ordinary matter of the involved WIMP particle.
This is the main experimental result of DAMA/NaI.

No other experiment has been realized so far able to exploit this model independent approach 
with similar exposed mass, sensitivity and control of the running parameters 
as DAMA/NaI.


\subsubsection{The corollary model dependent analyses searching for the nature of a candidate}

A corollary investigation can be pursued 
to investigate the nature and coupling
with ordinary matter of a possible candidate. In this case, a suitable energy and time
correlation analysis is necessary as well as the choice of a complete model  
framework. We remark that a model framework is identified
not only by the general
astrophysical, nuclear and particle physics assumptions, but also by the set
of values used for all the experimental and theoretical parameters needed in the calculations 
(for example WIMP local velocity, $v_0$, form factors and 
related parameters, quenching factors, halo model, coupling, etc., 
which are affected by relevant uncertainties)
\footnote{Note that
results given in terms of exclusion plots by experiments such as \cite{40,41,42} 
also depend on their own choice of nuclear,
particle and astrophysical assumptions and of experimental/theoretical parameters values; 
thus they have no generality at all.}. 

For simplicity, initially we have considered the particular case of purely
spin-independent (SI) coupled WIMP. In fact,  
often the spin-independent interaction
with ordinary matter is assumed to be dominant since e.g.
most of the used target-nuclei are practically
not sensitive to SD interactions as on the contrary
$^{23}$Na and $^{127}$I are and the theoretical calculations are even
more complex when including also this latter kind of interaction.
Moreover, only one set of possible parameters values has
been adopted \cite{21,23}. Then, this case has
been extended by considering some of the existing uncertainties 
on the astrophysical velocity distribution \cite{24,25} and the physical constraint
which arises from the upper limits on recoils measured by the 
same set-up \cite{25}. 
More recently an investigation on the effect induced on the result by considering
other possible and consistent halo models still for the particular case of purely SI
coupled WIMPs has also been carried out in ref. \cite{29}.

Moreover, some of the other possible scenarios
have also been considered such as extensions to the general case of WIMPs with
both spin-independent (SI) and spin-dependent (SD) coupling \cite{27} 
and to the case of WIMPs with preferred inelastic scattering \cite{28}.
In these latter cases, the effect of the uncertainties on some 
of the parameters has been included, but still only the simplified, approximate
and non-consistent isothermal halo model has been considered in the calculations.

Theoretical implications in terms
of a neutralino with dominant SI interaction
have been discussed e.g. in ref. \cite{43,44} for some theoretical model frameworks
and in terms of an heavy neutrino of the fourth family
in ref. \cite{45}.

\vspace{0.8cm}
\noindent{5.1.2.1 WIMPs with dominant SI interaction in given model frameworks}
\vspace{0.3cm}

A full energy and time correlation
analysis -- properly accounting for the physical constraint arising
from the measured upper limit on recoils \cite{26,30}
-- has been carried out in the framework of given
model for purely spin-independent coupled candidates with mass above 30 GeV
(this bound being inspired by the lower bound on the supersymmetric candidate,
as derived from the LEP data in the usually adopted supersymmetric schemes based on
GUT unification assumptions).
A standard
maximum likelihood method has been used.
Note that different model frameworks (see above)
vary the theoretical expectations and, therefore, the best fit values of 
cross section and mass (as well as the allowed region) also vary. 
In particular,
the inclusion of the uncertainties associated to the models and to
every parameter in the models themselves
as well as other possible scenarios largely 
enlarges the allowed region
as discussed e.g. in ref. \cite{24}
for the particular case of the astrophysical velocities  and offers
very large sets of best fit values.
We take this occasion to stress that ref. \cite{25} also accounts for 
several models; thus, for several 
sets of best fit values \footnote{
For example, for the particular model frameworks and assumptions of ref.
\cite{25}, where the WIMP local velocity, $v_0$, has been varied from 170 
km/s to 270 km/s
to account for its present uncertainty, we obtained the best fit values
$m_W = (72^{+18}_{-15})$ GeV and
$\xi \sigma_{SI} = (5.7 \pm 1.1) \cdot 10^{-6}$ pb for 
$v_0$ = 170 km/s and 
$m_W = (43^{+12}_{-9})$ GeV and
$\xi \sigma_{SI} = (5.4 \pm 1.0) \cdot 10^{-6}$ pb
for $v_0$ = 220 km/s. Here, $\xi$ is the WIMP local density
in 0.3 GeV cm$^{-3}$ unit,
$\sigma_{SI}$ is the point-like SI WIMP-nucleon
generalized cross section and $m_W$ is the WIMP mass.}.
Therefore, claims for contradiction in model dependent
comparisons 
by "choosing" a particular set of best fit values, given there as an example, are -- also in this respect --
arbitrary and substantially wrong.

As mentioned, more recently 
possible departures from the isothermal sphere model, which is the 
parameterisation usually adopted only in direct WIMP searches 
to describe the halo although approximate 
and non-consistent, have been investigated 
in a systematic way in ref. \cite{29} by some of us in collaboration with N. Fornengo and
S. Scopel. Modifications arising from various matter density profiles, 
effects due to anisotropies of the velocity dispersion tensor and rotation 
of the galactic halo have been specifically investigated. 
In particular, several halo models with potential and matter density 
having a spherical symmetry have been investigated as well as
halo models with spherical symmetry but anisotropic WIMP velocity distribution 
and halo models with axial symmetry (in these latter cases possible co-rotation or
counter-rotation of the dark halo has also been considered).
Also some triaxial models have been investigated.

\begin{figure}[!ht]
\vspace{-1.5cm}
\centering
\includegraphics[width=10.cm,angle=270]{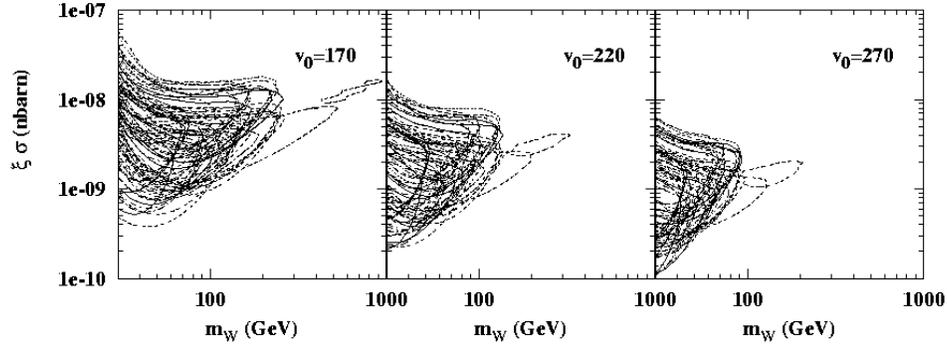}
\vspace{-2.6cm}
\caption{Superposition of the regions allowed at 3$\sigma$ C.L. in the given
model frameworks
by considering the velocity distribution of each one
of the halo models in ref. $^{28)}$.
Three of the possible values of $v_0$ are considered. Obviously 
a specific set of best fit values for the WIMP mass and
cross section corresponds to each
region. Note that inclusion of other existing
uncertainties will further enlarge the regions and increase the sets of best fit values.}
\label{fg:fig7}
\end{figure}

\begin{figure}[!ht]
\centering
\vspace{-0.5cm} 
\includegraphics[height=7.cm]{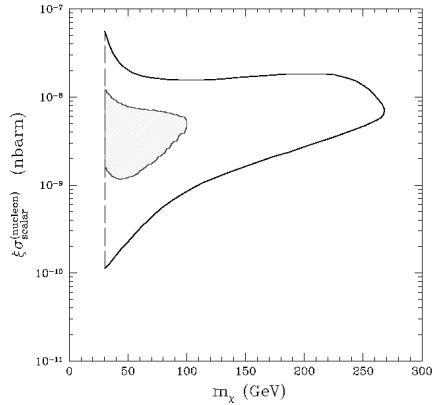}
\vspace{-0.9cm}
\caption{Region allowed at 3$\sigma$ C.L. given by the superposition of all the allowed
regions
obtained, in the given model frameworks, considering 
several possible non-rotating halo models $^{28)}$. 
Note that in these calculations only uncertainties on the halo model
have been considered; the inclusion of the other existing uncertainties would further enlarge 
it. It is evident that this cumulative region, as those given e.g. 
in ref. $^{23,24)}$, accounts for a large set of best fit values for 
WIMP mass and cross section.
The shaded region (which corresponds to the
particular case of the approximate and non-consistent isothermal sphere halo model
when assuming also $v_0 = 220 $ km/s and $\rho_0 = 0.3 $ GeV/cm$^{3}$ for the astrophysical, nuclear
and particle physics assumptions and fixed parameters of ref. $^{28)}$) is shown 
to point out only the effect due to the poor knowledge of the right halo model.}
\label{fg:fig8}   
\end{figure}

The global results of this analysis -- which includes also those presented in ref. 
\cite{25} -- are shown in Fig.
\ref{fg:fig7}, where the allowed regions obtained for the various considered halo models 
in given framework for three of the possible values of the local velocity,
$v_0$, above WIMP mass of 30 GeV are superimposed. Obviously different best fit values correspond to each one.   
The cumulative result, which gives a direct impact of the
effect induced on the region allowed in the considered scenario
only by the present poor knowledge of the right halo model, is shown 
in Fig. \ref{fg:fig8}.
This region is compared 
with the one obtained when considering, for a particular 
model framework (of the many possible), 
the approximate and non-consistent 
isothermal sphere halo model assuming in particular also $v_0 = 220 $ km/s e $\rho_0 = 0.3
$ GeV/cm$^{3}$. As one can see, 
the cumulative 3 $\sigma$ C.L. allowed region is extended up to $m_W
\simeq 270$ GeV with cross section on nucleon in the range 
$10^{-10} {\textrm{nbarn}} \leq \xi\sigma_{SI} \leq 6 \cdot 10^{-8}
{\textrm{nbarn}}$.
Maximal co-rotating and counter-rotating models can extend the allowed region
up to  $m_W \simeq 500-900$ GeV (see Fig. \ref{fg:fig7}).
We further stress that in this analysis no other uncertainty than the halo model
\footnote{We note that, although the large number of halo models considered in this analysis,
many other halo models are still possible, available and not yet considered here.}
has been considered; the proper inclusion 
of the other existing uncertainties will further extend the cumulative allowed region
and offer further sets of best fit values.
In particular, as already stressed also e.g. in ref. \cite{25}, in all the analyses 
cautiously the Helm SI form factor has been adopted, which is the most cautious one for 
Iodine. The use of other form factors enhancing the expected signal from Iodine 
recoils would yield
allowed regions corresponding to lower cross section values. The same would be if a
spin-dependent component different from zero would be introduced \cite{27} (see also next
section).

\vspace{0.3cm}
\noindent{\it 5.1.2.2 WIMPs with mixed coupling in given model frameworks}
\vspace{0.3cm}

Since the $^{23}$Na and $^{127}$I nuclei are sensitive
to both SI and SD couplings -- on the contrary e.g. of
$^{nat}$Ge and $^{nat}$Si which are sensitive mainly
to WIMPs with SI coupling (only 7.8 \% is non-zero spin isotope in
$^{nat}$Ge
and only 4.7\% of $^{29}$Si in $^{nat}$Si) --
the analysis of the data has been extended
considering the more general case of
a WIMP having not only a spin-independent,
but also a spin-dependent coupling different from zero
\footnote{We take this occasion to note that on the contrary of what is reported on
\cite{46},
ref. \cite{47} does not exclude at all a possible spin-dependent solution. In fact,
it refers only to the two particular cases (of the many possible)
for purely spin-dependent coupling: specifically the case where the effective coupling 
constant on proton is equal to
zero and the case where the effective coupling constant on neutron is equal to
zero. In addition, not only these two cases are handled in strongly model dependent
mode, but are even based on the wrong use of modulation amplitudes calculated for a particular purely
spin-independent case. Thus no restriction at all arises -- by the fact -- from ref. \cite{47} 
for any possible purely SD solution. In addition, the mixed case was not involved at all 
in that discussion.}.

In this case the free parameters are 
$\xi \sigma_{SI}$, $\xi \sigma_{SD}$ and $m_W$
for each given $\theta$ value, where $\sigma_{SD}$ is the point-like SD WIMP
cross section on nucleon and $tg\theta$ is the ratio
between the effective SD coupling constants on neutrons, $a_n$,
and on proton, $a_p$; therefore, $\theta$ can assume all the values between
0 and $\pi$ depending on the type of SD coupling.

Note that the results in ref. \cite{27} have been obtained by
considering there only the approximate non-consistent isothermal 
sphere model to describe the galactic halo and a maxwellian
WIMP velocity distribution with inclusion of some    
uncertainties on $v_0$, 
on the nuclear radius and on the nuclear surface thickness parameter in the used Helm SI form factor,
on the $b$ parameter in the used used Ressel e al. SD form factor
and on the quenching factors. As an additional source of uncertainty we mention that 
an universal formulation is not possible for the SD form factor,
thus other formulations than the one adopted here are
possible and can be considered with evident
implications on the obtained results. The same is for the used spin factor.

\begin{figure}[ht]
\centering
\vspace{0.5cm}
\includegraphics[height=8.cm]{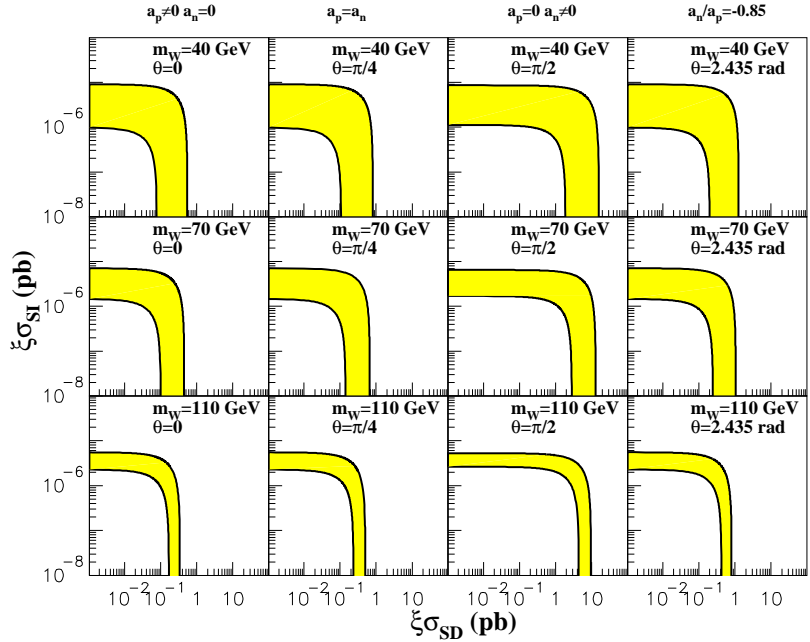}
\caption{A mixed SI/SD case: example of slices
of the region allowed at 3 $\sigma$ C.L.
in the ($\xi \sigma_{SI}$, $\xi \sigma_{SD}$, $m_W$) space for some
$m_W$ and $\theta$
values in the model frameworks considered in ref. $^{26)}$.
Only four particular couplings are reported here for simplicity:
i) $\theta$ = 0;
ii) $\theta$ = $\pi/4$
iii) $\theta$ = $\pi/2$; iv)
$\theta$ = 2.435 rad. Note that e.g. Ge experiments are sensitive mainly only
to SI coupling and, therefore, cannot explore most of the DAMA allowed regions in
this scenario; the same is in most cases also for $^{nat}$Xe since the odd-spin isotopes have 
the neutron as unpaired nucleon.
These allowed regions would be further enlarged by taking into account
the uncertainties existing on the halo models
and related parameters, on the SD form factor (which has a not an universal formulation)
and on some other experimental and theoretical parameters and assumptions.}
\label{fg:fig9}
\end{figure}

For simplicity, Fig. \ref{fg:fig9} shows slices for some
$m_W$ of the region allowed -- in the model frameworks of ref. \cite{27} -- at 3 $\sigma$ C.L.
in the ($\xi \sigma_{SI}$, $\xi \sigma_{SD}$, $m_W$) space
for four particular couplings:
i)  $\theta$ = 0 ($a_n$ =0 and $a_p \ne$ 0 or $|a_p| >> |a_n|$);
ii) $\theta = \pi/4$ ($a_p = a_n$);
iii)  $\theta$ = $\pi/2$ ($a_n \ne$ 0 and $a_p$ = 0 or  $|a_n| >> |a_p|$);
iv) $\theta$ = 2.435 rad ($ \frac {a_n} {a_p}$ = -0.85, pure Z$^0$
coupling).
The case $a_p = - a_n$ is nearly similar to the case iv).

As already pointed out, when the SD contribution goes to zero
(y axis in Fig. \ref{fg:fig9}), an interval not
compatible with zero is obtained for $\xi\sigma_{SI}$.
Similarly, when the SI contribution goes to zero
(x axis in Fig. \ref{fg:fig9}), finite values for the SD cross section are
obtained.
Large regions are allowed for mixed configurations in the considered model frameworks
also for $\xi\sigma_{SI} \lsim 10^{-5}$ pb and
$\xi\sigma_{SD} \lsim 1$ pb; only in the particular case of
$\theta = \frac {\pi} {2}$ (that is $a_p = 0$ and $a_n \ne 0$)
$\xi\sigma_{SD}$ can increase up to $\simeq$ 10 pb, since
the $^{23}$Na and $^{127}$I nuclei have the proton as odd
nucleon. Moreover, in ref. \cite{27} we have also pointed out that:
i) finite values can be allowed for $\xi\sigma_{SD}$
even when $\xi\sigma_{SI} \simeq 3 \cdot 10^{-6}$ pb as in the region
allowed
in the pure SI scenarios considered above;
ii) regions not compatible with zero
in the $\xi\sigma_{SD}$ versus $m_W$ plane are allowed
even when $\xi\sigma_{SI}$ values much lower than those allowed in
the dominant SI scenarios previously summarized are considered; iii)
best fit values with both $\xi\sigma_{SI}$ and
$\xi\sigma_{SD}$ different from zero are present for some $m_W$ and
$\theta$ pairs; the related confidence level ranges between
$\simeq$ 3 $\sigma$ and $\simeq$ 4 $\sigma$ \cite{27}.

Further investigations are in progress on these
model dependent analyses to account for other known parameters
uncertainties
and for possible different model assumptions.
In fact, as mentioned, when including the uncertainties on the halo models
and their parameters, on the SD form factor, on the spin factor
and on some other experimental and theoretical parameters, the allowed volume 
in the space ($\xi \sigma_{SI}$, $\xi \sigma_{SD}$, $m_W$) for each $\theta$ value
would be further enlarged as well as the number of obtained sets of best fit values 
for the cross sections and the WIMP mass.

In conclusion, this analysis has shown that the DAMA data
of the four annual cycles, analysed in terms
of WIMP annual modulation signature,
can also be compatible with a mixed scenario
where both $\xi\sigma_{SI}$ and $\xi\sigma_{SD}$ are
different from zero. 

\vspace{0.3cm}
\noindent{\it 5.1.2.3 Inelastic Dark matter}
\vspace{0.3cm}

It has been suggested in ref. \cite{48} that the
observed annual modulation effect could be induced by possible
inelastic Dark Matter: relic particles that prefer to scatter inelastically
off of nuclei. The inelastic Dark Matter
could arise from a massive complex scalar split into two approximately
degenerate real scalars or from a Dirac fermion split into two
approximately degenerate Majorana fermions, namely $\chi_+$ and $\chi_-$,
with a $\delta$ mass splitting. In particular, a specific
model featuring a real component of the sneutrino,
in which the mass splitting naturally arises, has been given in ref.
\cite{48}.
It has been shown that for the $\chi_-$ inelastic scattering
on target nuclei a kinematical constraint exists which favours
heavy nuclei (such as $^{127}$I) with respect to
lighter ones (such as e.g. $^{nat}$Ge) as target-detectors media.
In fact, $\chi_{-}$ can only inelastically scatter
by transitioning to $\chi_{+}$ (slightly heavier state than $\chi_{-}$)
and this process can occur
only if the $\chi_{-}$ velocity is larger than
$v_{thr} = \sqrt{\frac{2\delta}{m_{WN}}}$
where $m_{WN}$ is the WIMP-nucleus reduced mass ($c=1$).
This kinematical constraint becomes increasingly severe
as the nucleus mass, $m_N$, is decreased \cite{48}.
Moreover, this model scenario gives rise -- with respect to the
case of WIMP elastically scattering -- to an enhanced
modulated component, $S_m$, with respect to the unmodulated one, $S_0$,
and to largely different behaviours with energy for
$S_0$ and $S_m$ (both show a higher mean value) \cite{48}.

A dedicated energy and time
correlation analysis of the DAMA annual modulation data has been carried out
\cite{28} handling
aspects other than the interaction type
as in ref. \cite{27}.
\begin{figure}[htb]
\centering
\vskip 0.2cm
\includegraphics[height=9.cm]{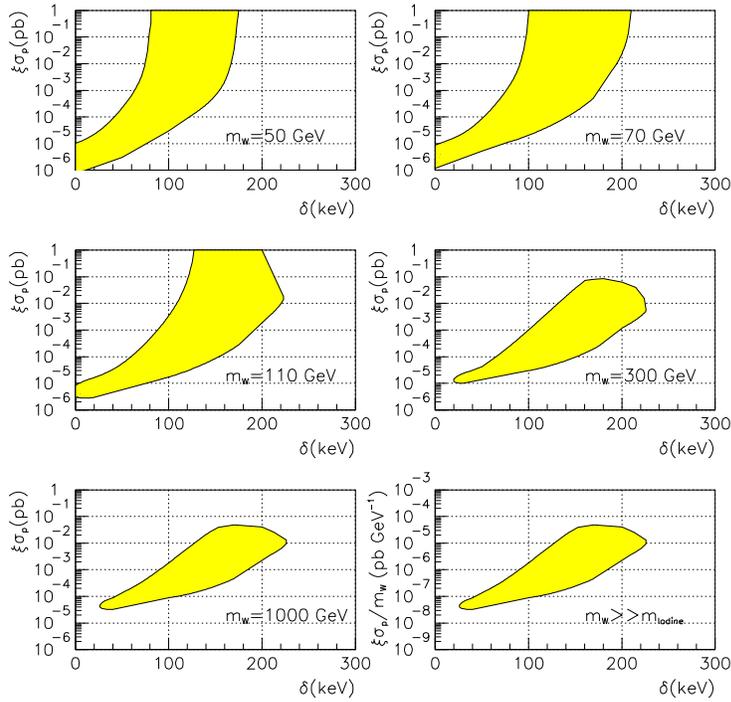}
\caption{An inelastic case: slices at fixed WIMP masses of the volume
allowed at 3 $\sigma$ C.L. in the space ($\xi \sigma_p$, $\delta$, $m_W$)
obtained for the model frameworks considered in ref. $^{27)}$; some of 
the uncertainties on used parameters have been included. 
Note that e.g. Ge experiments 
cannot explore most of the DAMA allowed regions in 
this scenario.
These allowed regions would be further enlarged by taking into account
the uncertainties existing on the halo models
and their parameters and on some other experimental and theoretical parameters.}
\label{fg:fig10}
\vskip -0.2cm
\end{figure}
In this scenario of Dark Matter with inelastic scattering an allowed volume in the space
($\xi \sigma_p$, $\delta$, $m_W$) is obtained \cite{28}. For simplicity,
Fig. \ref{fg:fig10} shows slices of such an allowed volume at some given 
masses (3 $\sigma$ C.L.) for the model frameworks considered in ref. \cite{28}.
It can be noted that when $m_W \gg m_N$,
the expected differential energy spectrum is trivially dependent on $m_W$
and in particular it is proportional to the ratio between $\xi \sigma_p$
and $m_W$; this particular case is summarized in the last
plot of Fig. \ref{fg:fig10}.
The allowed regions have been obtained
-- as in the previous cases -- by the superposition
of those obtained when varying the values of some of the used 
parameters according to ref. \cite{28}.
Of course, each set of parameters' values 
gives rise to a different expectation, thus to different
"most likely" values.
Just as an example we mention that when fixing
the other parameters as in ref. \cite{27},
the "most likely" values for a WIMP mass of 70 GeV
are: i) $\xi \sigma_p$ = $2.5 \times
10^{-2}$ pb and $\delta = 115$ keV when $v_0 = 170$ km/s,
ii) $\xi \sigma_p$ = $6.3 \times
10^{-4}$ pb and $\delta = 122$ keV when $v_0 = 220$ km/s; they are in 
$\delta$ region where e.g. Ge and Si experiments are disfavoured.

Finally, we note again that the allowed regions are further enlarged
when properly including the uncertainties on the halo models,
on the experimental and theoretical parameters and on the other assumptions. 

\subsubsection{Comparison with some model dependent results}

\vspace{0.3cm}
\noindent{\it 5.1.3.1 ... from direct searches}
\vspace{0.3cm}

As mentioned above no other experiment directly comparable with the model independent 
DAMA/NaI result on WIMPs in the galactic halo is available at present; thus, 
claims for contradiction 
are intrinsically arbitrary/wrong.

Only few experiments \cite{40,41,42},
which use different target nuclei and
different methodological approaches,
have released extremely poor selected statistics
quoting an exclusion plot in a given particular model framework.

Table \ref{tb:modepcomp1} shortly summarizes some main
items.
\begin{table}[!ht]
\caption{Features of the DAMA/NaI results on the WIMP annual modulation signature 
during the first four annual cycles (57986 kg $\times$ day exposure) 
$^{20,22,23,24,25,26,27,28)}$
and those of refs. $^{39,40,41)}$.} 
\begin{center}
\scriptsize
\begin{tabular}{|c|c|c|c|c|}
\hline \hline
  &  &  & & \\
  & DAMA/NaI  & CDMS-I & Edelweiss-I & Zeplin-I \\
  &  &  & & \\
\hline\hline
  &  &  & &\\
Signature    & annual  & None  & None & None   \\
  & modulation &  & & \\
\hline
  &  &  &   & \\
Target-nuclei & $^{23}$Na, $^{127}$I & $^{nat}$Ge &  $^{nat}$Ge &  $^{nat}$Xe \\
\hline
  &  &  &   & \\
Technique    & well known &  poorly & poorly & critical optical  \\
  &  & experienced  & experienced & liquid/gas interface \\
  &  &  & &  in this realization\\
\hline
  &  &  & & \\
Target mass &  $\simeq 100$ kg & 0.5 kg & 0.32 kg & $\simeq 3$ kg\\
\hline
  &  &  & & \\
Exposure     & 57986 kg $\times$ day & 15.8 kg $\times$ day  
                  & 8.2 kg $\times$ day  & 280 kg $\times$ day\\
\hline
  &  &  & & \\
Depth of the  & 1400 m  & 10 m  & 1700 m & 1100 m \\
experimental site &  &  & & \\
\hline
  &  &  & & \\
Software energy & 2 keV  & 10 keV  & 20 keV & 2 keV  \\ 
threshold & (5.5 -- 7.5 p.e./keV) &  & & (but: $\sigma/E =100$\%   \\
(electron   &  &  & & mostly   \\
equivalent) &  &  & & 1 p.e./keV; \cite{49})\\
 &  &  & & (2.5 p.e./keV \\
 &  &  & & for 16 days; \cite{42}) \\
\hline
  &  &  & & \\
Quenching       & Measured & Assumed  = 1 & Assumed = 1 & Measured  \\
factor &  &  & &  \\
\hline
  &  &  & & \\
Measured event & $\simeq 1$ cpd/kg/keV  & $\simeq 60$ cpd/kg/keV & 
2500 events & $\simeq 100$ cpd/kg/keV \\
rate in low &   &(10$^5$ events)  & total &  \\
energy range & &  & & \\
\hline
  &  &  & & \\
Claimed events  & & 23 in Ge, 4 in Si, & 0 &  $\simeq$ 20-50 cpd/kg/keV \\
after rejection & &  4 multiple evts in Ge &  &  after rejection and \\
procedures      & & + MonteCarlo on    & &  ?? after standard PSD \\
                     & &  neutron flux  & & \cite{42,49} \\
\hline
  &  &  & & \\
Events satisfying & modulation           & & & \\ 
the signature     & amplitude      & & & \\ 
in DAMA/NaI       & integrated over the  & & & \\ 
                  & given exposure & & & \\ 
                  & $\simeq 2000$ events & & & \\ 
\hline
                & & from few down        & from few down        & \\ 
Expected number & & to zero depending    & to zero depending    & depends on   \\
of events from  & & on the models        & on the models        & the models \\ 
DAMA/NaI effect & & (and on quenching    & (and on quenching    &(even zero) \\
                & & factor)              & factor)              & \\
\hline\hline
\end{tabular}
\end{center}
\label{tb:modepcomp1}
\end{table}
As it can be seen, the mentioned exclusion plots 
are based on a huge data selection (releasing
typically extremely poor exposures with    
respect to generally long data taking and, in some cases, to several used detectors). Moreover, 
their counting rate is very high and few/zero events are claimed after applying
several strong and hardly safe rejection procedures 
(involving several orders of magnitude; see Table \ref{tb:modepcomp1}).
These rejection procedures are also poorly described and, often,
not completely quantified.  
Moreover, most efficiencies and physical quantities entering in the interpretation of
the claimed selected events have never been discussed in the needed 
details. As an example, we mention the case 
of the quenching factor of the recoil target nuclei in
the whole bulk material for the bolometer cases, which is arbitrarily 
assumed to be 1 \footnote{In fact, no direct measurement performed with neutron 
sources or generators has been reported up to now, 
although several bolometers have been irradiated with
neutrons along the past decade.
For the sake of completeness, we remind that
a measurement of the response of a TeO$_2$ bolometer to surface $^{224}$Ra
recoiling nuclei has been reported in ref. \cite{Cuore};
this measurement, although its importance, does not represent a determination of the
quenching factor for a bolometer since neither the target-nuclei nor the whole
bulk of the detector are involved. Moreover, these values cannot 
of course be extended to whatever kind of bolometer.}, 
implying a substantially arbitrarily assumed
energy scale and energy threshold (and consequently arbitrary 
exclusion plot).
Further uncertainties are present when, as in
ref. \cite{40}, a neutron background modeling and 
subtraction is pursued in addition.

As regards in particular the Zeplin-I result of ref. \cite{42,49},
a very low energy threshold is claimed (2 keV), although the 
light response is  very poor: between $\simeq$ 1 ph.e./keV \cite{49}
(for most of the time)
and $\simeq$ 2.5 ph.e./keV (claimed for 16 days) \cite{42} 
\footnote{For comparison we remind that the data of the DAMA/LXe
set-up, which has a similar light response, are analysed by using the much more realistic
and safer software energy threshold of 13 keV \cite{9}.}.
Moreover, a strong data filtering is applied to
the high level of measured counting rate ($\simeq$ 100 cpd/kg/keV at low energy, 
which is nearly two orders of magnitude larger than the DAMA NaI(Tl) background in the 
same energy region)
by hardware vetoes, by fiducial volume cuts and,
largely, by applying down to few keV a standard pulse shape discrimination procedure,
although the LXe scintillation pulse
profiles (pulse decay time $<$ 30 ns) 
are quite similar even to noise events in the lower energy bins and in spite of the
poor light response. Quantitative information on experimental quantities related to the used
procedures has not yet been given \cite{42,49}.

In addition to the experimental aspects, these experiments 
-- which cannot perform any model independent comparison with the 
result of DAMA/NaI -- generally uncorrectly/partially 
quote the results of published quests for a purely SI coupled 
candidate in given model frameworks and ignore the 
effects of the relevant uncertainties in the astrophysical, nuclear 
and particle physics assumptions and experimental/theoretical parameters
values taken by each experiments as well as the interpretation of the 
DAMA/NaI model independent effect in terms of candidates 
with other kind of coupling. 
All that can significantly vary the result of any comparison
(even when assuming as correct the evaluation of the selected number
of events, the energy scale and the energy threshold determinations 
given in refs. \cite{40,41,42,49} and the used efficiencies, etc.).
In addition, there exist scenarios to which Na and I are sensitive
and other nuclei, such as e.g. 
$^{nat}$Ge, $^{nat}$Si and $^{nat}$Xe, are not. 

In conclusion: 
\begin{enumerate}
\item  
no other experiment, whose result can be directly comparable in a model independent
   way with that of DAMA/NaI, is available so far.
\item
as regards in particular CDMS-I, EDELWEISS-I and Zeplin-I, e.g.:

   i)   they are insensitive to the model independent WIMP annual modulation
        signature exploited by DAMA/NaI; 
   ii)  they use different methodological approaches, which do not allow any 
        model independent comparison and they have different sensitivities to 
        WIMPs; in particular, the number of counts they could expect on the 
        basis of the model independent DAMA/NaI result varies from few to zero 
        events depending on the models, on the assumptions and on the 
        theoretical/experimental parameters' values adopted in the calculations;  
   iii)	they make neither correct nor complete comparisons with the DAMA/NaI experimental
        result;   
   iv)  they use extremely poor statistics; 
   v)   they reduce their huge measured counting rate 
        of orders of magnitude by various rejection procedures
        claiming for very optimistic rejection powers;
   vi)  their energy scale determination and/or
        energy threshold appear questionable (in the first two cases 
        because of the quenching factors values
        and in the second because of the poor number of photoelectrons/keV);
    etc. 

\end{enumerate}

\vspace{0.3cm}
\noindent{\it 5.1.3.2 ... from indirect searches}
\vspace{0.3cm}

As regards results from indirect searches,
model dependent analyses investigating possible up-going muons arising from
WIMP annihilation in celestial bodies have been carried out 
by large experiments deep underground such as e.g. Macro and Superkamiokande.
The comparison is strongly model dependent;
anyhow it has already been shown in ref. \cite{43} that
even in the simple model frameworks considered in ref. \cite{25}
the model dependent limit on up-going muon flux obtained by
Macro (and thus that by Superkamiokande, which is only
slightly more stringent) could cut only part of the allowed 
configurations in MSSM.

On the other hand, as regards antimatter searches carried out outside the 
atmosphere, we remind the analysis of the HEAT balloon-borne experiment 
performed in ref. \cite{50}, where 
an excess of positrons with energy $\simeq 5 - 20$ GeV 
has been found and has been interpreted in
terms of WIMP annihilation.
We also mention the analysis of ref. \cite{51} which already suggests
the presence of a $\gamma$ excess from the center of the Galaxy
in the EGRET data  \cite{52} which matches with a possible
WIMP annihilation in the galactic halo and is not in 
conflict with the DAMA/NaI model independent result.

We take this occasion to stress, however, that the specific 
parameters of a WIMP candidate 
(mass and cross sections), which 
can be derived from the indirect searches, critically depend
on several assumptions used in the calculations such as the estimation 
of the background, the halo model, the amount of WIMP in the galactic dark halo, 
the annihilation channels, the transport of charged particle to the Earth, etc.; thus,
they have the same relative meaning as those obtained in the quest for 
a candidate in direct search approach as previously described.

\vspace{0.3cm}
\noindent{\it 5.1.3.3 Conclusions}
\vspace{0.3cm}

 In conclusion, no model independent comparison with the DAMA/NaI effect is available. 
Only few model dependent approaches have been used in direct searches 
to claim for a particular model
dependent comparison, which appears in addition -- as mentioned above --  
neither based on solid procedures nor fully correct nor complete. 
On the other hand, the indirect search approaches, which can also offer  
only model dependent comparisons, are not in contradiction or 
in substantial agreement with the DAMA/NaI observed effect.

Thus, the interest in the further available DAMA/NaI data is increased
and the analysis of the cumulative data of the seven annual cycles
will be released in near future.

\section{The new DAMA/LIBRA set-up}

After the completion of the data taking of the DAMA/NaI set-up, 
the procedures to install
the new LIBRA set-up have been carried out.
In particular, improvements have been realized in the
experimental site, in the Cu box,
in the shield and in the available detectors.

The LIBRA set-up consisting of $\simeq 250$ kg of radiopure NaI(Tl)
is made of 25 detectors, 9.70 kg each one.
The new detectors have been realized thanks to a second generation R\&D
with Crismatec company, by exploiting in particular new radiopurification
techniques of the NaI and TlI powders. In the framework of this
R\&D new materials have also been selected, prototypes have been built and tested 
and a devoted protocol has been fixed.

\begin{figure}[!htpb]
\centering
\vskip -0.2cm
\includegraphics[height=5.cm]{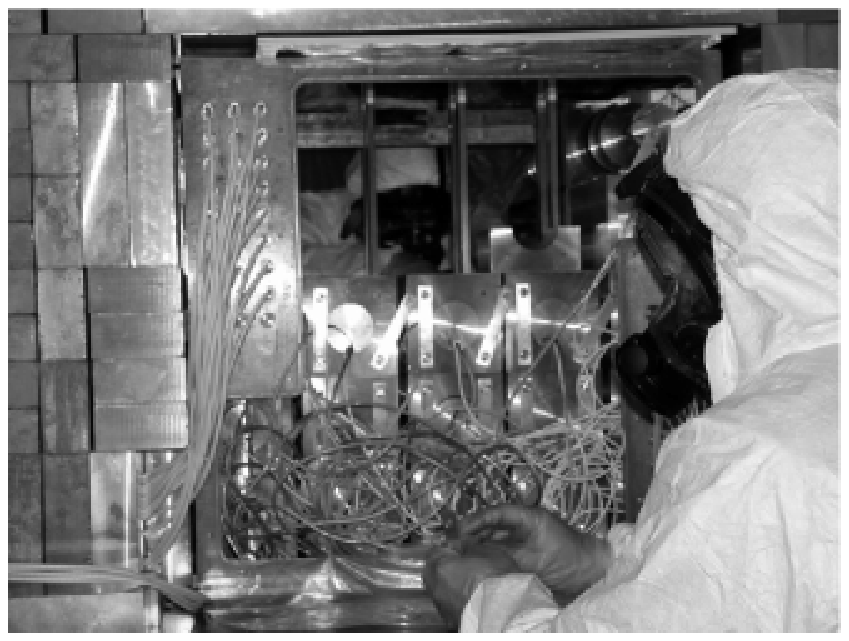}
\includegraphics[height=5.cm]{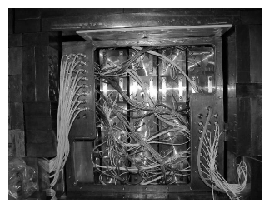}
\caption{Left picture: during the detectors installation
in HP Nitrogen atmosphere.
Right picture: view at end of the detectors installation.
All the used materials have been deeply selected for radiopurity 
(see for example the cables with teflon envelop).}
\label{fg:fig11}
\end{figure}

The full dismounting of the $\simeq 100$ kg NaI(Tl)
set-up, the improvements mentioned above and the installation of all
the detectors including a new PMTs' shield have been completed at end 2002.
All the related procedures 
have been performed in HP Nitrogen atmosphere by using
special masks connected to air bottles
to avoid that the inner part of the Cu box, the detectors, the new PMTs'
shield, etc.
would be in contact with environmental air, that is to
reduce at most possible surfaces' contamination by environmental  
Radon. 
Some pictures taken during the installation of the 
LIBRA detectors are shown in Fig. \ref{fg:fig11}. 
Before this installation, all the Cu parts have been 
chemically etched following a devoted protocol and maintained in HP
Nitrogen atmosphere until the installation.

LIBRA will offer a relevant competitiveness e.g. because of its:
i)    high sensitivity;
ii)   standard and well defined operating procedures; 
iii)  well known technology; 
iv)   proved possibility of an effective control of the experimental 
      conditions during several years of running;
v)    high duty cycle;
vi)   possibility to deeply investigate the WIMP model independent signature;
vii)  sensitivity to both SI and SD couplings;
viii) favoured sensitivity in some of the possible particle and astrophysical models;
ix)    high benefits/cost.

The main aim of this new LIBRA set-up is to further investigate the
WIMP component in the galactic halo with increased sensitivity thanks to the larger 
exposed mass and to the higher
overall radiopurity.
Moreover, when applicable, it offers also pulse shape discrimination capability
and the possibility to achieve competitive results in the searches for 
other rare processes.
For example, it would reach in a relatively short time a sensitivity
of $\simeq 3 \times 10^{27}$ y for possible Pauli-exclusion-principle violating processes,
a sensitivity of $\simeq 10^{24} - 10^{25}$ y for possible charge-non-conserving processes
in $^{23}$Na and in $^{127}$I (depending on the counting rate in the energy region of interest),
a sensitivity of $\simeq 10^{27}$ y for the nucleon and di-nucleon decay into invisible
channels, a sensitivity of $\simeq 10^{-10}$ GeV$^{-1}$ on the axion-photon coupling constant
from the investigation of solar axions; in addition, it could explore mass for exotic Dark Matter
candidates, such as e.g. the SIMPs, up to 
$\simeq 10^{17}$ GeV and neutral nuclearites flux of $\simeq 5 \times 10^{-12}$ s$^{-1}$cm$^{-2}$sr$^{-1}$.

\section{Conclusion}

In this paper recent results achieved by the DAMA
experiment at the Gran Sasso National Laboratory of I.N.F.N. have been summarized.
In particular,
DAMA/NaI has been a pioneer experiment running at LNGS for several years and 
investigating as first the WIMP annual modulation signature 
with suitable sensitivity and control of the running parameters. It has pointed out the presence of a modulation
satisfying the several peculiarities of a WIMP induced effect
and the absence of any possible systematic effects or side reactions able to mimic it,
reaching a significant model independent evidence.
As a corollary result, it has also pointed out the complexity of the quest for a WIMP
candidate because e.g. of the present poor knowledge on the many astrophysical,
nuclear and particle physics aspects and on its nature.

After the completion of the data taking of the $\simeq 100$ kg NaI(Tl)
set-up (on July 2002; full statistics of 107731 kg $\cdot$ d in progress 
to be released), as a result of the
continuous efforts toward the creation of ultimate radiopure 
set-ups, the new DAMA/LIBRA has been realized and installed.
Further R\&D efforts for ultimate radiopurification of NaI(Tl) 
detectors are also starting again.

\end{document}